\documentclass[letterpaper]{article} 
\usepackage{aaai25}  
\usepackage{times}  
\usepackage{helvet}  
\usepackage{courier}  
\usepackage[hyphens]{url}  
\usepackage{graphicx} 
\usepackage{natbib}  
\usepackage{caption} 
\usepackage{algorithm}
\usepackage{algorithmic}

\usepackage{amssymb}
\usepackage{amsmath}
\usepackage{booktabs}
\usepackage{multirow}
\usepackage{subcaption}
\usepackage{newfloat}
\usepackage{listings}
\usepackage[table]{xcolor}

\newcommand{\htwoo}{\ensuremath{\mathrm{H}_2\mathrm{O}}}
\newcommand{\behtwo}{\ensuremath{\mathrm{BeH}_2}}
\newcommand{\htwo}{\ensuremath{\mathrm{H}_2}}

\DeclareCaptionStyle{ruled}{labelfont=normalfont,labelsep=colon,strut=off} 
\floatstyle{ruled}
\newfloat{listing}{tb}{lst}{}
\floatname{listing}{Listing}
\title{BenchRL-QAS: Benchmarking Reinforcement Learning Algorithms for Quantum Architecture Search}
\author{
	Azhar Ikhtiarudin\equalcontrib\textsuperscript{\rm 1},
	Aditi Das\equalcontrib\textsuperscript{\rm 2},
	Param Thakkar\textsuperscript{\rm 3},
	Akash Kundu\textsuperscript{\rm 4}\footnote{Corresponding Author}
}
\affiliations{
	\textsuperscript{\rm 1} Computational Materials Design \& Quantum Engineering Group, Bandung Institute of Technology, Bandung, Indonesia\\
	\textsuperscript{\rm 2}Department of Physics, National Institute of Technology, Karnataka, India\\
	\textsuperscript{\rm 3}Veermata Jijabai Technological Institute, Mumbai, India\\
	\textsuperscript{\rm 4}QTF Centre of Excellence, Department of Physics, University of Helsinki, Helsinki, Finland\\
	azharikhtiarudin@gmail.com, aditi.das1601@gmail.com, akash.kundu@helsinki.fi
}

\begin{document}

\maketitle

\begin{abstract}
 
 We present BenchRL-QAS, a unified benchmarking framework for reinforcement learning (RL) in quantum architecture search (QAS) across a spectrum of variational quantum algorithm tasks on 2- to 8-qubit systems. Our study systematically evaluates 9 different RL agents, including both value-based and policy-gradient methods, on quantum problems such as variational eigensolver, quantum state diagonalization, variational quantum classification (VQC), and state preparation, under both noiseless and noisy execution settings. To ensure fair comparison, we propose a weighted ranking metric that integrates accuracy, circuit depth, gate count, and training time. Results demonstrate that no single RL method dominates universally, the performance dependents on task type, qubit count, and noise conditions providing strong evidence of {no free lunch principle} in RL-QAS. As a byproduct we observe that a carefully chosen RL algorithm in RL-based VQC outperforms baseline VQCs. {BenchRL-QAS establishes the most extensive benchmark for RL-based QAS to date}, codes and experimental made publicly available for reproducibility and future advances.
 
\end{abstract}

%
\begin{links}
   \link{Code}{https://github.com/azhar-ikhtiarudin/bench-rlqas}
\end{links}

\section{Introduction}

\begin{table*}[t]
	\centering
	\fontsize{9}{9}\selectfont
	\begin{tabular}{l|cccc|cccc|cccc|cccc}
		\toprule
		\multicolumn{1}{l}{} 
		& \multicolumn{4}{c}{State preparation$^*$}  
		& \multicolumn{4}{c}{VQSD}
		& \multicolumn{4}{c}{VQE ($\texttt{H}_2\texttt{O}$)} 
		& \multicolumn{4}{c}{VQC} \\
		\cmidrule(lr){2-5} \cmidrule(lr){6-9} \cmidrule(lr){10-13} \cmidrule(lr){14-17}
		\multicolumn{1}{l}{RL algorithms} 
		& \multicolumn{1}{c}{C} & \multicolumn{1}{c}{D} & \multicolumn{1}{c}{A} & \multicolumn{1}{c}{T}
		& \multicolumn{1}{c}{C} & \multicolumn{1}{c}{D} & \multicolumn{1}{c}{A} & \multicolumn{1}{c}{T}
		& \multicolumn{1}{c}{C} & \multicolumn{1}{c}{D} & \multicolumn{1}{c}{A} & \multicolumn{1}{c}{T}
		& \multicolumn{1}{c}{C} & \multicolumn{1}{c}{D} & \multicolumn{1}{c}{A} & \multicolumn{1}{c}{T} \\
		\midrule
		A2C~\cite{sutton1998reinforcement}  &            &           & \checkmark & \checkmark  &  &  &  &  &  &  &  &  & \checkmark & \checkmark & & \\
		A3C~\cite{mnih2016asynchronous}  & \cellcolor{green!30}\checkmark & \cellcolor{green!30}\checkmark  & \cellcolor{green!30}\checkmark &  &  &  &  &  &  &  &  & \checkmark &  & \checkmark & & \checkmark \\
		DQN~\cite{mnih2013playingatarideepreinforcement} &            &           & \checkmark &  & \cellcolor{green!30}\checkmark & \cellcolor{green!30}\checkmark & \cellcolor{green!30}\checkmark & \cellcolor{green!30}\checkmark & \checkmark &  &  & \checkmark  & \checkmark  & & \\
		DDQN~\cite{van2016deep}  & \cellcolor{green!30}\checkmark & \cellcolor{green!30}\checkmark & \cellcolor{green!30}\checkmark & \cellcolor{green!30}\checkmark  & \cellcolor{green!30}\checkmark &  & \cellcolor{green!30}\checkmark & \cellcolor{green!30}\checkmark & \cellcolor{green!30}\checkmark & \cellcolor{green!30}\checkmark & & \cellcolor{green!30}\checkmark & \cellcolor{green!30}\checkmark  & \cellcolor{green!30}\checkmark  & \\
		DQN\_rank~\cite{schaul2016prioritizedexperiencereplay}  &            &           & \checkmark &  & \checkmark  &  & \checkmark &            &           & \checkmark &  \checkmark & & \checkmark  &  & \\
		DQN\_per~\cite{schaul2016prioritizedexperiencereplay}  &            &           & \checkmark &  &  &  &  &  & \cellcolor{green!30}\checkmark &  &  \cellcolor{green!30}\checkmark & \cellcolor{green!30}\checkmark & & \checkmark  & \\
		Dueling DQN~\cite{wang2016dueling} &            &           & \checkmark &  &  &  & \checkmark & \checkmark & \checkmark & \checkmark & & & \cellcolor{green!30}\checkmark  & \cellcolor{green!30}\checkmark  & \cellcolor{green!30}\checkmark & \cellcolor{green!30}\checkmark \\
		PPO~\cite{schulman2017proximal} &  &   & \checkmark & \checkmark  &  &  &  & \checkmark & \checkmark &  &  & \checkmark &  &  &  &  \\
		TPPO~\cite{wang2020trulyproximalpolicyoptimization} &  &  & \checkmark & \checkmark & \cellcolor{green!30}\checkmark & \cellcolor{green!30}\checkmark &  & \cellcolor{green!30}\checkmark & \checkmark & \checkmark &  & &  &  &  &  \\
		\bottomrule
	\end{tabular}
	\caption{\textbf{Weighted ranking performance (see main text) of RL algorithms is reported for four quantum tasks}: state preparation, VQSD, VQE ($\texttt{H}_2\texttt{O}$), and VQC. Metrics include smallest circuit (C), depth (D), accuracy (A), and episode time (T), averaged over 5–10 different neural network initializations. A (\checkmark) marks the best result, showing the task and criterion dependence of RL methods for quantum circuit design. "$*$" indicates a non-parameterized action space. For completeness, we also benchmarked VQE ($\texttt{BeH}_2$); results showed similar trends across RL methods and are omitted here for compactness. DQN\_per uses prioritized replay~\cite{schaul2016prioritizedexperiencereplay}; DQN\_rank applies rank-based prioritization within DQN\_per.}
	\label{tab:noiseless}
\end{table*}

Quantum computing offers new possibilities for tackling computational challenges that are beyond the reach of classical systems. Yet, the promise of quantum advantage remains difficult to realize, largely due to the practical limitations of current noisy intermediate-scale quantum (NISQ) devices~\cite{preskill2018quantum}. Constraints such as limited qubit counts, hardware connectivity issues, and high error rates collectively limit both the scale and complexity of quantum circuits that can be executed. As a consequence, many theoretically powerful quantum algorithms are not yet implementable on today's hardware~\cite{monz2016realization}, underscoring the need for new algorithms specifically tailored to current quantum technologies.

To address these limitations, hybrid quantum-classical strategies have become increasingly popular in the NISQ era. Among these, variational quantum algorithms (VQAs) stand out as a leading framework for utilizing NISQ hardware efficiently~\cite{peruzzo2014variational, cerezo2021variational, bharti2022noisy}. VQAs leverage classical optimization to tune the parameters of parameterized quantum circuits (PQCs), minimizing a task-specific cost function, often related to the expectation value of a Hamiltonian. The success of this approach depends crucially on the choice of \textit{ansatz}, which defines the PQC structure. Traditionally, these circuit layouts are selected in advance, guided either by hardware constraints~\cite{kandala2017hardware} or informed by heuristic, problem-specific criteria~\cite{peruzzo2014variational}. However, such manually designed circuits frequently face an inherent tension between expressiveness and noise resilience, limiting the scalability of VQAs~\cite{cerezo2021variational, bharti2022noisy, LaroccaBPReview2025}.

Recent advances have focused on quantum architecture search (QAS)~\cite{zhang2022differentiable}, aiming to automate the discovery of optimal PQC structures from a broad set of quantum gates. QAS methodologies adaptively construct circuit architectures that are tailored both to the computational task and to hardware restrictions, searching systematically for gate sequences and placements that maximize performance. Among various QAS techniques, reinforcement learning (RL) has become a particularly promising tool for navigating the large and discrete design space of quantum circuits~\cite{kuo2021quantum, fosel2021quantum, ostaszewski2021reinforcement}. In this context, RL agents build PQCs step-by-step, making sequential gate selections and improving their policies based on feedback from quantum performance metrics. Although there have been successful demonstrations of RL-based QAS for circuits with up to 20 qubits~\cite{kundu2025tensorrl}, a broad and systematic understanding of which RL algorithms are most effective for different quantum optimization tasks remains lacking. Closing this gap is critical for advancing both QAS methodologies and the broader goal of practical quantum advantage with near-term hardware.

A wide variety of RL agents have recently been deployed in QAS for VQAs (see Tab.~\ref{tab:summary_RL-QAS}). While these studies highlight the potential of RL techniques for quantum circuit construction, the field is still missing a thorough and comprehensive benchmarking of these algorithms. Most prior evaluations focus on only a narrow subset of RL agents, usually in isolated settings and without standardized evaluation criteria. This makes it challenging to determine which algorithms consistently achieve desirable outcomes, be it minimizing circuit depth, reducing 1- and 2-qubit gate usage, or optimizing solution accuracy on quantum processing units (QPUs). The absence of extensive benchmarking keeps open the question of the best-suited RL algorithm for different QAS objectives. Notably, recent works such as~\cite{zhu2023quantum} and~\cite{altmann2024challengesreinforcementlearningquantum} have compared only a few RL variants, and then solely on non-parameterized action spaces, highlighting the need for systematic benchmarking of parameterized action spaces, particularly in the NISQ context. Beyond variational quantum algorithms, the RL framework has potential applications across a wide range of scientific and engineering disciplines. RL has been used in quantum control~\cite{bukov2018reinforcement} and error correction~\cite{nautrup2019optimizing, olle2024simultaneous}. In quantum chemistry, RL-assisted ansatz synthesis accelerates ground-state energy estimation of molecules while reducing circuit depth and gate count~\cite{ostaszewski2021reinforcement}. Moreover, RL is applied to higher energy models~\cite{biswas2024safe, kundu2025improving}.

In this work, we address this critical gap by presenting a unified benchmark encompassing a broad range of RL algorithms-including both value-based and policy-gradient methods, across several quantum optimization tasks. In particular, we investigate tasks including variational quantum state diagonalization (VQSD), variational quantum eigensolver (VQE), variational quantum classification (VQC), and state preparation, covering system sizes from 2- to 8-qubit. Our benchmark systematically evaluates nine distinct RL agents on each problem. To ensure statistical robustness, each experiment is repeated 5–10 times with independent neural network initializations, resulting in a dataset of $325$ separate quantum optimization problem instances. By evaluating all RL approaches under consistent conditions, we provide a comprehensive account of the strengths and limitations of each algorithm for QAS. The highlights of our results are presented in Tab.~\ref{tab:noiseless} (noiseless setting) and Tab.~\ref{tab:noise} (noisy settings), where we employ a \textit{weighted ranking performance estimator}, fully described later in the paper, to identify the best RL agent for each problem type.


\subsection{Contributions}
\begin{itemize}
	\item We present \textbf{BenchRL-QAS}, a platform for systematic benchmarking of RL methods in quantum architecture search (QAS) across diverse variational quantum algorithms (VQAs).
	\item We provide the largest benchmark to date of RL algorithms for QAS, evaluating parameterized and non-parameterized action spaces under both noiseless and noisy settings. Results show no RL algorithm is universally optimal, demonstrating the \textit{no free lunch}~\cite{wolpert1997no} principle in QAS.
	\item We offer a comparative analysis of RL effectiveness, identifying strengths of specific algorithms and giving recommendations to guide future RL-QAS research.
\end{itemize}

\section{Related Work}

\paragraph{Quantum Architecture Search (QAS)}
\begin{table}[t]
	\centering
	\fontsize{9}{9}\selectfont
	\label{table:rl-algorithms-in-qas}
	\renewcommand{\arraystretch}{1.4}
	\begin{tabular}{@{}p{4cm}p{3cm}@{}} 
		\toprule
		Related works & RL algorithms \\
		\midrule \cite{ostaszewski2021reinforcement, patel2024curriculum, kundu2024enhancing, kundu2025tensorrl, kundu2025reinforcementlearninggadget} & DDQN \\
		\cite{ye2021quantum} & PPR-DQN \\
		\cite{olle2025scalingautomateddiscoveryquantumgadget, fodera2024reinforcement} & PPO \\
		\cite{kuo2021quantum, fosel2021quantum} & A2C, PPO \\
		\cite{lockwood2021optimizing} & TD3, SAC, PPO \\
		\cite{patel2024reinforcement} & REINFORCE \\
		\cite{altmann2024challengesreinforcementlearningquantum} & A2C, PPO, SAC, TD3 \\
		\cite{zhu2023quantum} & A2C, PPO, TRPO \\
		\cite{dutta2025qasqtn} & A2C, PPO, DDQN, TD3 \\
		\bf {BenchRL-QAS (this work)} & \bf A2C, A3C, DQN, DQN\_per, DQN\_rank, Dueling DQN, DDQN, PPO, TPPO \\
		\bottomrule
	\end{tabular}
		\caption{\textbf{Summary of RL algorithms used in QAS}. In our work, DQN\_per refers to DQN with prioritized experience replay~\cite{schaul2016prioritizedexperiencereplay} and DQN\_rank is the rank based prioritization in DQN\_per.}
	\label{tab:summary_RL-QAS}
\end{table}
One of the early efforts to automate quantum circuit design involves using search heuristics based on evolutionary genetics to generate more efficient and novel quantum circuits \cite{williams1998automated}. Despite their effectiveness in exploring large problem spaces, evolutionary algorithms often yield problem-specific circuits \cite{rattew2019domain, chivilikhin2020mog, huang2022robust}.
Another approach is differentiable quantum architecture search \cite{zhang2022differentiable, chen2025differentiable}, which is inspired by the DARTS method in neural architecture search \cite{liu2018darts}.
Grimsley et al.~\cite{grimsley2019adaptive} introduced an adaptive approach for constructing quantum circuits, known as ADAPT-VQE, which has since been further improved \cite{ramoa2024reducing}.
Other techniques have also been explored, including generative models \cite{nakaji2024generative, F_rrutter_2024}, Bayesian optimization \cite{duong2022quantumbayesianoptimization, PhysRevABayesianOptimization}, and Monte Carlo tree search \cite{wang2023automatedqasmontecarlo}.
Recent work has proposed QAS methods driven by circuit topology \cite{su2025topologydrivenquantumarchitecturesearch} and by landscape fluctuation analysis \cite{zhu2025scalablequantumarchitecturesearch}, as well as training-free approaches that rank circuit feature maps via fast proxy metrics \cite{gujju2025quprofsevolutionarytrainingfreeapproach}.

\paragraph{Reinforcement Learning for QAS} Numerous studies have successfully applied reinforcement learning (RL) techniques to optimize quantum circuit architectures. These applications include variational quantum eigensolver (VQE) optimization \cite{ostaszewski2021reinforcement}, the construction of multi-qubit maximally entangled states \cite{ye2021quantum}, and the optimization of quantum machine learning models \cite{lockwood2021optimizing}. Building on these efforts, Patel et al.~\cite{patel2024curriculum} introduced curriculum RL combined with advanced pruning techniques, which were later utilized by Dutta et al.~\cite{dutta2025qasqtn}, while Tang et al.~\cite{tang2024alpharouter} integrated RL with Monte Carlo tree search and Ruiz et al.~\cite{ruiz2025quantum} employed tensor decomposition methods. RL-based variational quantum algorithms (VQAs) have also been applied to solve combinatorial optimization problems \cite{patel2024reinforcement}, quantum state diagonalization tasks \cite{kundu2024enhancing}, and Maximum Cut problems \cite{fodera2024reinforcement, fosel2021quantum}. More recently, RL-based quantum architecture search (RL-QAS) has been further enhanced through the incorporation of tensor network techniques \cite{kundu2025tensorrl} and composite gate constructions (gadgets) \cite{olle2025scalingautomateddiscoveryquantumgadget, kundu2025reinforcementlearninggadget}. A summary of the reinforcement learning algorithms used in previous works is provided in Table~\ref{tab:summary_RL-QAS}.

\section{BenchRL-QAS}
The BenchRL-QAS framework provides a systematic and reproducible platform for benchmarking reinforcement learning (BenchRL) algorithms in quantum architecture search (QAS) across a diverse set of variational quantum algorithms. It addresses the need for standardized evaluation in quantum circuit design, supporting key challenges such as variational quantum state diagonalization (VQSD), variational quantum eigensolver (VQE), variational quantum classification (VQC), and GHZ state preparation. Each task is formulated as a circuit optimization problem, enabling RL agents to discover efficient circuit architectures based on metrics like accuracy, depth, and gate count. The framework is defined in Algorithm \ref{algo:BenchRL-QAS}. In the following we define the RL-state, action space, reward function, illegal actions to accelerate the performance of the agent and a weighted ranking metric to evaluate RL agents.

\paragraph{RL-State} The state of the agent is defined by a tensor-based encoding of the current quantum circuit~\cite{patel2024curriculum}. This encoding scheme captures both the structural arrangement of quantum gates, depth, their parameter values, and current achieved accuracy providing a compact yet expressive representation of the ansatz and its performance. The ansatz is expressed as a tensor of dimension $[D_\text{max} \times ((N + 3) \times N )]$, where $N$ number of qubits and $D_\text{max}$ is the considered maximum depth of the ansatz.

\paragraph{The Reward Function}
The reward function \( R \) in the BenchRL-QAS framework (for VQE, VQSD and VQC problems), is given by~\cite{ostaszewski2021reinforcement}:

\begin{equation}
	\label{eq:reward-function}
	R = 
	\begin{cases}
		5 & \text{if } C_t \leq \zeta, \\
		-5 & \text{if } t \geq D_\text{max} \text{ and } C_t \geq \zeta, \\
		\max\left( \dfrac{C_{t-1} - C_t}{C_{t-1} - E_{\min}}, -1 \right) & \text{otherwise}.
	\end{cases}
\end{equation}

Here \( C_t(\vec{\theta}) \) is the cost function at step \( t \), and \( \zeta \) is a predefined convergence threshold i.e. the ansatz accuracy and is treated as a hyperparameter. The \( \zeta \) is problem specific and discussed in the experimental settings. For the state preparation task the reward is described by~\cite{kundu2024kanqas}:
\begin{equation}
	R = \begin{cases}
		\mathcal{R}, & \text{if } F(s_t) \geq 0.98 \\
		F(s_t),      & \text{otherwise}
	\end{cases}
	\label{eq:reward-function-state}
\end{equation}
R is a hyperparameter reward with $R\gg F(s_t)$ and $F(s_t)$ is the fidelity of state at step $t$.

\paragraph{Action Space}
The action space in BenchRL-QAS is tailored to the quantum task, supporting both parameterized and non-parameterized circuit construction. For parameterized tasks such as VQE, VQSD, and VQC, it is hybrid: each action specifies a gate type (\texttt{RX}, \texttt{RY}, \texttt{RZ}, \texttt{CX}), its target qubit(s), and, for parameterized gates, a continuous parameter value. This enables the RL agent to simultaneously optimize both the discrete circuit structure and continuous gate parameters. In contrast, for non-parameterized tasks like GHZ state preparation, the action space is limited to a discrete set of gates (\texttt{CX}, \texttt{X}, \texttt{Y}, \texttt{Z}, \texttt{H}, \texttt{T}), where actions involve selecting and placing gates on specific qubits, without continuous parameters. This design allows the framework to benchmark RL agents across a wide range of quantum circuit design problems.
\paragraph{Illegal Actions}
In BenchRL-QAS, we focus on enforcing only the two most critical illegal action constraints during quantum architecture search. Specifically, an action $a$ is considered illegal in state $s$ if it satisfies either of the following analytical conditions:
\begin{align*}
	&\text{(Redundancy)} \quad G_{m,q} = G_{m-1,q} \\
	&\text{(\texttt{CX} repetition)} \quad G_{m,(q_1,q_2)} = \texttt{CX} \ \wedge \ G_{m-1,(q_1,q_2)} = \texttt{CX}
\end{align*}
where $G_{m,q}$ denotes the gate applied to qubit $q$ at moment $m$, and $G_{m,(q_1,q_2)}$ denotes a 2-qubit gate between control $q_1$ and target $q_2$. Other potential constraints, such as hardware connectivity, circuit depth, or parameter validity, are not enforced in our implementation. To ensure the agent avoids these illegal actions during training, we assign $Q(a, s) = -\infty$ whenever either of the above conditions is met, effectively masking such actions from the policy optimization.

\begin{algorithm}[t!]
	\caption{BenchRL-QAS}
	\begin{algorithmic}[1]
		\REQUIRE Quantum tasks $\mathcal{T}$, RL algorithms $\mathcal{A}$, encoding scheme, illegal action handling, curriculum (if used)
		\FOR{task $t$ in $\mathcal{T}$}
		\STATE Initialize environment $\mathcal{E}_t$ for $t$
		\FOR{RL algorithm $a$ in $\mathcal{A}$}
		\STATE Init agent $\mathcal{R}_a$ and empty quantum circuit $C$
		\WHILE{not converged}
		\STATE $s \leftarrow$ \texttt{EncodeState}$(C)$
		\STATE $u \leftarrow \mathcal{R}_a.\texttt{SelectAction}(s)$
		\IF{\texttt{IsIllegalAction}$(u)$}
		\STATE Penalize or mask, update agent; \textbf{continue}
		\ENDIF
		\STATE Update $C \leftarrow \texttt{ApplyAction}(C,u)$
		\STATE $r \leftarrow \mathcal{E}_t.\texttt{Evaluate}(C)$
		\STATE $s' \leftarrow \texttt{EncodeState}(C)$
		\STATE $\mathcal{R}_a.\texttt{Update}(s,u,r,s')$
		\ENDWHILE
		\STATE Log metrics for $(t,a)$
		\ENDFOR
		\STATE [Curriculum] Increase difficulty, if enabled
		\ENDFOR
		\STATE Aggregate and compare results
	\end{algorithmic}
	\label{algo:BenchRL-QAS}
\end{algorithm}

\paragraph{Weighted Ranking Approach for Agent Evaluation}
To objectively compare RL algorithms across multiple criteria, we use a \textit{weighted ranking} scheme~\cite{ayan2023comprehensive}. The key metrics, average circuit error ($E$), number of gates ($G$), circuit depth ($D$), and time per episode ($T$) are each normalized to $[0,1]$ (lower is better): $X_\text{norm} = \frac{X - X_\text{min}}{X_\text{max} - X_\text{min}}$, where $X$ is the metric value for a given algorithm, and $X_\text{min}$ and $X_\text{max}$ are the minimum and maximum values of that metric across all algorithms. The composite score for each algorithm is then computed as
\begin{equation}
	S = w_E E_\text{norm} + w_G G_\text{norm} + w_DD_\text{norm} + w_T T_\text{norm}.
	\label{eq:weighted_ranking_metric}
\end{equation}
Lower $S$ indicates better overall performance. Algorithms are ranked by ascending $S$, ensuring that accuracy is the most influential criterion in the final ranking, while still incorporating resource and efficiency considerations. We explicitly choose the weights $[w_E, w_G, w_D, w_T] = [0.5, 0.2, 0.2, 0.1]$ for noiseless results in Tab.~\ref{tab:noiseless} and $[0.6, 0.1, 0.3, 0.0]$ for noisy in Tab.~\ref{tab:noise}, placing the greatest emphasis on problem accuracy (circuit error), while still accounting for circuit size, depth, and computational efficiency.

\section{Experimental Settings}~\label{app:exp-details}
Our experimental settings span a range of quantum circuit design tasks, each presenting a distinct challenge for RL-based quantum architecture search. VQSD requires agents to diagonalize arbitrary quantum states without Hamiltonian dependence; VQE targets ground state energy approximation for specific Hamiltonians; VQC trains parameterized circuits for supervised learning on synthetic data; and GHZ state preparation constructs non-parameterized circuits to generate maximally entangled states essential for quantum information. By casting these as circuit optimization problems, BenchRL-QAS enables systematic evaluation of RL agents on accuracy, circuit depth, and gate count, providing a unified and versatile benchmarking platform. Throughout the paper BenchRL-QAS utilizes vanilla curriculum~\cite{ostaszewski2021reinforcement}. Moreover we utilize $9$ different RL-agents 
and benchmark the performance of these agents. In the following we utilize neural network consists of $L$ layers, each containing $1000$ neurons, a batch size of $1000$, and a replay memory of $20000$ transitions. The learning rate is set to $3\times10^{-4}$ using the \texttt{ADAM}~\cite{kingma2014adam}, with no dropout, and the target network is updated every 500 steps. Exploration is controlled by an epsilon-greedy policy with epsilon decaying from $1.0$ to a minimum of $0.05$ at a rate of $0.99995$ per step, and a discount factor gamma of $0.88$ is used. Circuit parameters are optimized using \texttt{COBYLA} with a maximum 500 iterations. we detail the algorithms and their agent-environment specifications. Furthermore, specifically for the A3C agent we utilize 3 workers for all optimization problems. All the results are obtained utilizing \texttt{AMD Rome 7H12} CPU based on \texttt{AMD Zen 2} architecture and an
\texttt{Nvidia Ampere A100} GPU. We utilized 2 CPUs with maximum 4GB and maximum 40 GB GPU memory.

\subsection{RL-VQSD}

Variational quantum state diagonalization (VQSD) is a type of variational quantum algorithm that aims to find a unitary transformation which diagonalizes a quantum state in the computational basis \cite{larose2019variational}. More specifically, given a quantum state $\rho$, the method optimizes parameters $\vec\theta$ of a parameterized unitary $U(\vec\theta)$ such that:
\begin{equation}
	\rho' = U(\vec\theta_{\text{opt}})\, \rho\, U(\vec\theta_{\text{opt}})^\dagger = \rho_{\text{diag}},
\end{equation}
where $\rho_{\text{diag}}$ is the diagonal form of $\rho$ in its eigenbasis. Compared to quantum principal component analysis \cite{lloyd2014quantum}, which promises exponential speedup but requires many qubits and deep circuits, VQSD is more suitable for current noisy quantum devices due to its shallow circuits and reduced hardware demands. VQSD has applications in fidelity estimation of quantum states \cite{cerezo2020variationalquantumfidelity}, and quantum device certification \cite{kundu2022variational}. A central challenge in implementing VQSD lies in designing a hardware-efficient and scalable ansatz. To address this challenge, reinforcement learning has recently been introduced as a tool to automate and optimize the construction of ansatz circuits for variational quantum algorithms (RL-VQSD) \cite{kundu2024enhancing}.

Under the BenchRL-QAS framework we consider several 2-qubit Haar random mixed quantum states and diagonalize them utilizing the VQSD algorithm. The agent-environment setup consists of maximum depth $D_\text{max}=40$, with the agent aiming to minimize a cost function below a threshold $\zeta=5\times10^{-2}$ in reward (see Eq.~\ref{eq:reward-function}) with $L=4$ layers of neural network.

\subsection{RL-VQE}

The variational quantum eigensolver (VQE) is designed to find the eigenstates and eigenvalues of a given Hamiltonian, with particular emphasis on estimating the ground state energy. VQE operates by minimizing a cost function defined as:
\begin{equation}
	\label{eq:vqe-cost-function}
	C(\vec{\theta}) \equiv E(\vec{\theta}) = \langle \psi(\vec{\theta}) | \hat{H}_q | \psi(\vec{\theta}) \rangle,
\end{equation}
where \( \hat{H}_q \) is the qubit Hamiltonian and \( |\psi(\vec{\theta})\rangle \) is a parameterized wavefunction. 
The qubit Hamiltonian \( \hat{H}_q \) is derived from the molecular system, which includes information such as molecular geometry and atomic charges. This leads to a fermionic Hamiltonian expressed in the second quantization formalism. The fermionic Hamiltonian is then mapped to a qubit representation using techniques such as the Jordan-Wigner \cite{fradkin1989jordan}, Bravyi-Kitaev \cite{bravyi2002fermionic}, and the qubit tapering \cite{bravyi2017taperingqubitssimulatefermionic}. 
VQE is a leading method in quantum chemistry and materials science for estimating molecular ground state energies, enabling insights into system properties, reaction prediction, and materials design~\cite{peruzzo2014variational, tilly2022variational, Fedorov2021}. Unlike quantum phase estimation~\cite{kitaev1995quantum}, which demands deep, complex circuits impractical for today’s hardware, VQE is well-suited to current quantum devices. However, traditional ansatzes like UCCSD and ADAPT-VQE~\cite{grimsley2019adaptive} often result in circuits too deep for near-term quantum processors, while hardware-efficient ansatzes face scalability and trainability challenges. To overcome these limitations, reinforcement learning-based approaches~\cite{ostaszewski2021reinforcement, patel2024curriculum} have been developed to automate and scale the construction of efficient VQE ansatzes.

In BenchRL-QAS the environments utilize 4-$\htwo$, 6-$\behtwo$, and 8-$\htwoo$ with $D_\text{max}$, $40$, $70$, and $250$ steps each mapped using the Jordan-Wigner transformation. The reward function is similar to that defined in Eq.~\ref{eq:reward-function}, but the cost function at each step is obtained by Eq.~\ref{eq:vqe-cost-function}. The main goal of the agent is to achieve chemical accuracy $\zeta=1.6\times10^{-3}$. The $L$ is of 3-, 4-, and 5- layers for 4-, 6-, and 8-qubit problem. The circuit optimization is performed using \texttt{COBYLA} with 100 (4-qubit), 200 (6-qubit), or 500 (8-qubit) iterations.

\subsection{RL-VQC}
\begin{table}[b]
	\centering
	
	\fontsize{9}{9}\selectfont
	\renewcommand{\arraystretch}{1.2}
	\begin{tabular}{@{}lccc@{}}
		\toprule
		Method & Train accuracy  & Test accuracy \\
		\midrule
		\textbf{Hardware efficient ansatz}\\
		\quad 2 layers &$90$\%& $85$\%&\\
		\quad 3 layers &$90$\%& $85$\%&\\
		\quad 4 layers &$92$\%& $92$\%&\\
		\textbf{\cite{du2022quantum}} \\
		\quad W = 1,  T = 10& N.A.& $50\% - 60\%$ &\\
		\quad W = 1,  T = 400&N.A.& $>90\%$ &\\
		\quad W = 5,  T = 400&N.A.& $>90\% $ &\\
		\cellcolor{green!30}\textbf{RL-VQC (This work)} & \cellcolor{green!30}\textbf{99.996}\% & \cellcolor{green!30}\textbf{99.991}\%\\
		\bottomrule
	\end{tabular}
	\caption{\textbf{Training and testing accuracy comparison:} RL-VQC with DQN and rank-based prioritized replay achieves the highest accuracy, outperforming hardware-efficient ansatz and net-based methods~\cite{du2022quantum} in the noiseless scenario without retraining.
	}
	\label{tab:RL-VQC_performance}
\end{table}
Variational quantum classifier (VQC) is a hybrid quantum-classical algorithm for supervised learning problems such as classification \cite{gil2019variational}. Like other variational quantum algorithms, VQC consists of a parameterized quantum circuit used in conjunction with a classical optimizer to learn from labeled data. For a given input $x=(x_{1},x_{2})\epsilon\mathbb{R}^2$, the quantum circuit initially encodes the data using single qubit rotations $R_{y}(x_{i}\pi )$. The encoded state is subsequently processed through a trainable ansatz comprising further rotation gates and optional entangling gates such as \texttt{CX} or \texttt{CZ}. The circuit ends with measurements on all qubits, and the outcomes are used to calculate a prediction-dependent cost function. The behavior of the circuit is based on the trainable parameters $\vec{\theta}=(\theta_{1},...,\theta_{l})$  that are learned by a classical algorithm (usually gradient descent) to minimize a quadratic cost function:
\begin{equation}
	\label{VQC_cost_func}
	C = \frac{1}{2n} \sum_{x} | y(x; \vec\theta) - a(x) |^2,
\end{equation}
where $y(x;\vec{\theta})$ is the circuit's output for input $x$ and a(x) is the true label. The model continues to improve iteratively to minimize this cost. After the VQC outputs measurement results interpreted as class labels, these predicted labels are compared against the true labels to compute accuracy. VQC have demonstrated better accuracy and noise robustness over classical neural networks in accelerator physics \cite{Yin2025} but the performance of VQC is highly dependent on diligent ansatz design, circuit depth, and task-dependent encoding schemes.
\begin{figure*}[t!]
	\centering
	\begin{subfigure}[b]{0.7\textwidth}
		\includegraphics[width=1\linewidth]{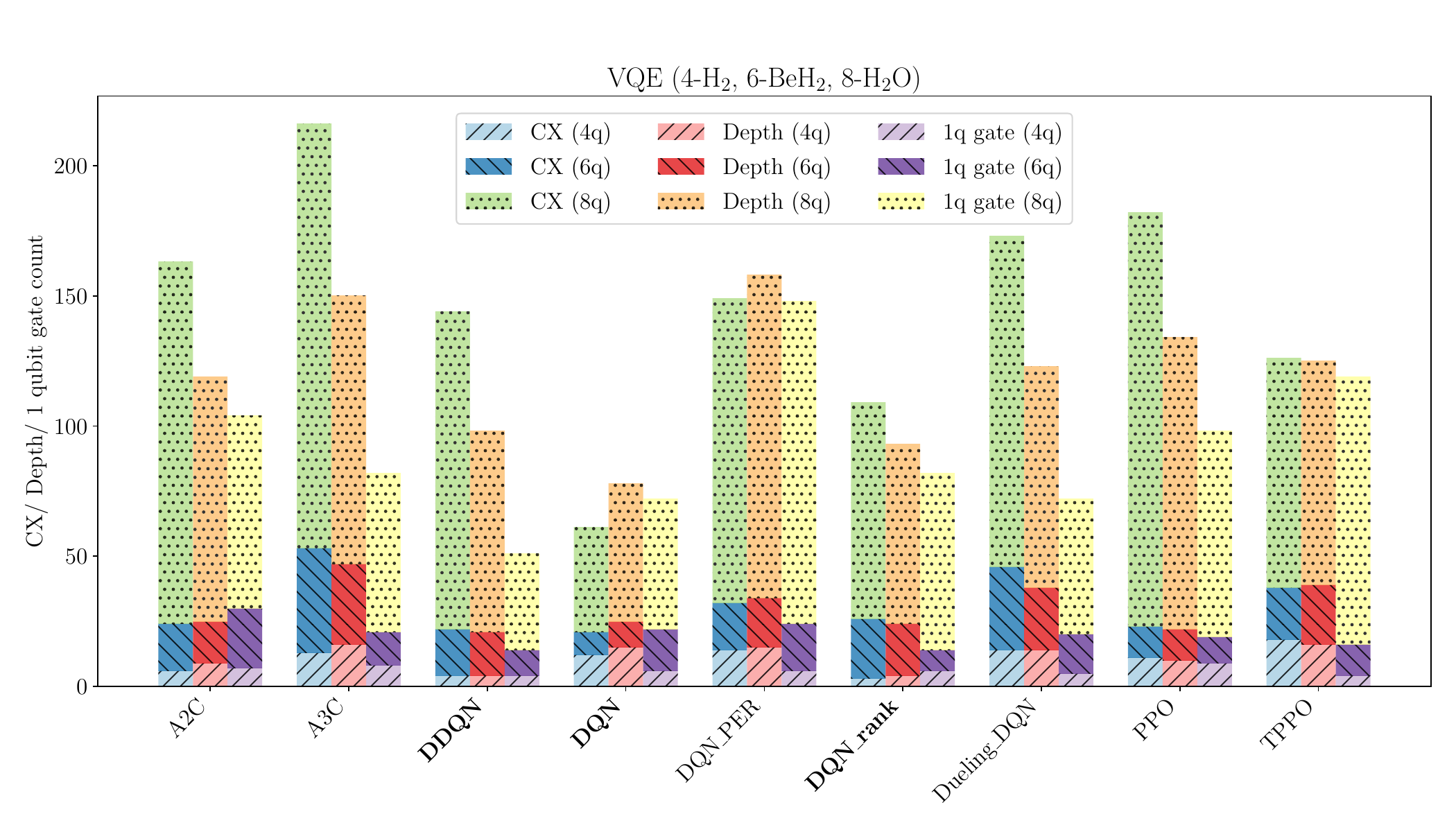}
	\end{subfigure}
	\begin{subfigure}[b]{0.7\textwidth}
		\includegraphics[width=1\linewidth]{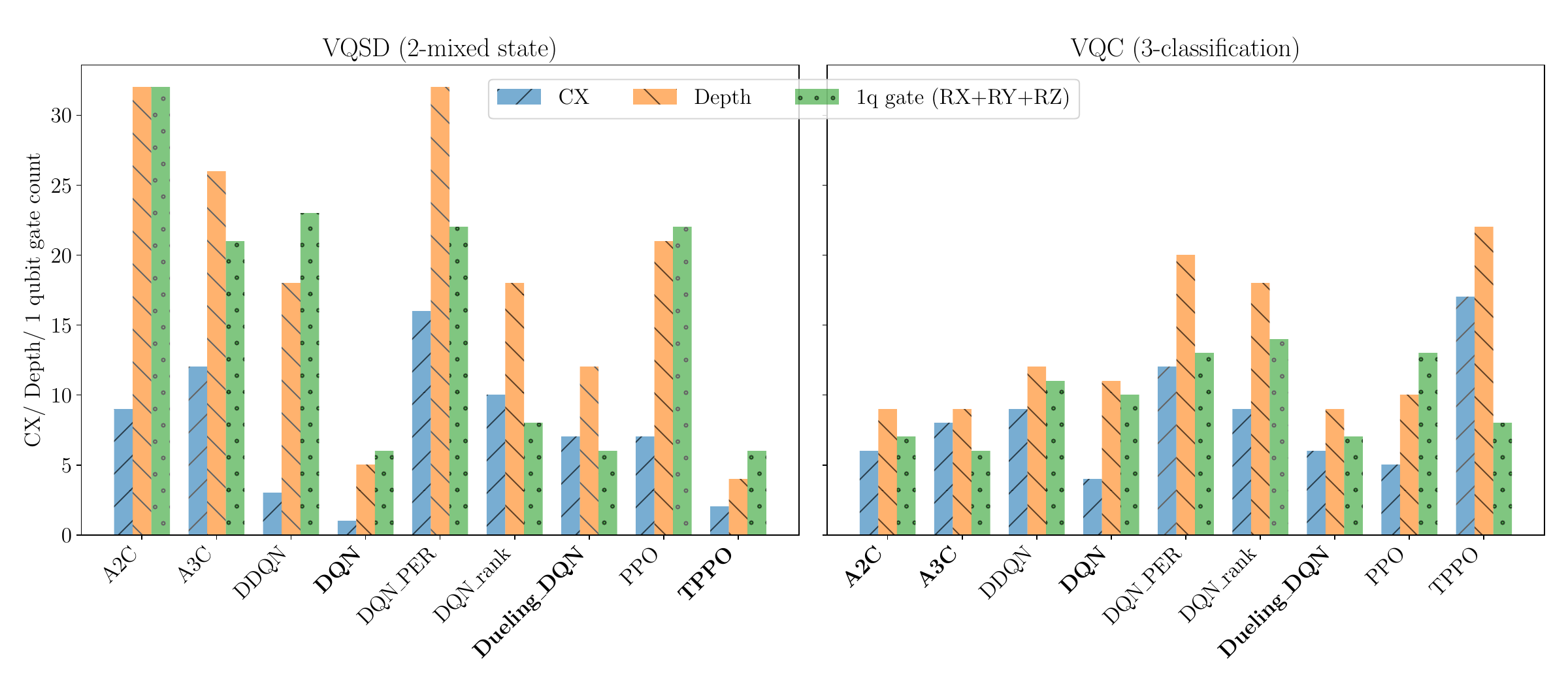}
	\end{subfigure}
	\caption[]{\textbf{Benchmark of circuit depth, CX, and single-qubit gate counts for RL-based quantum architecture search (QAS) across VQSD (2-qubit), VQC (3-qubit), and VQE (4-\htwo, 6-\behtwo and 8-\htwoo) in noiseless scenario}. Our RL approach delivers circuits for 6-\behtwo with errors over three orders of magnitude lower and gate counts less than half those of the recent TF-QAS method, establishing one of the new standards for QAS.}
	\label{fig:noiseless_benchmark}
\end{figure*}

To tackle this, we incorporate RL to automate the design of quantum circuit for VQC namely RL-VQC. We consider 3-qubit synthetic binary classification data with a predefined training error threshold $\zeta=0.2$ i.e. achieving at least 80\% accuracy in the training loss $\mathcal{L}_\text{training}$ within a maximum allowable depth $D_{max}=25$ and a neural net of $L=3$ layers. In the reward in Eq.\ref{eq:reward-function}, the cost function is evaluated according to Eq. \ref{VQC_cost_func}. The quantum circuit parameters are optimized using \texttt{COBYLA} with 1000 iterations per step. Under {BenchRL-QAS} we benchmark RL-VQC for $9$ different agents. The preliminary results in Tab.~\ref{tab:RL-VQC_performance} show that the DQN\_rank has the optimal performance, with {99.996\% training} and {99.991\% test accuracy}, outperforming both the hardware-efficient ansatz (HEA) , as well as net-based approaches~\cite{du2022quantum}. The HEA used here is a layered variational circuit of iterated single-qubit rotations \texttt{RY} and nearest-neighbor \texttt{CNOT} gates applied after a data-dependent \texttt{RY} feature encoding. A broader benchmark is provided in results section.

\subsection{RL State Preparation}
Beyond the VQAs, we also utilize the BenchRL-QAS quantum state preparation, which in this case preparing the Greenberger–Horne–Zeilinger (GHZ) state, defined as:
$|GHZ\rangle = \frac{1}{\sqrt{2}}(|000\rangle + |111\rangle)$,
with non-parameterized gateset (\texttt{CX}, \texttt{X}, \texttt{Y}, \texttt{Z}, \texttt{H}, \texttt{T}). The aim of BenchRL-QAS is to construct a quantum circuit that reproduces the $|GHZ\rangle$ state, the closer the RL outcome is to the $|GHZ\rangle$ state the higher the reward to the agent, following the reward in Eq.~\ref{eq:reward-function-state}.
\begin{table*}[t!]
	\centering
	
	\fontsize{9}{9}\selectfont
	\begin{tabular}{l|ccc|ccc|ccc}
		\hline
		\multirow{2}{*}{{RL algorithm}} & \multicolumn{3}{c|}{4-\htwo} & \multicolumn{3}{c|}{6-\behtwo} & \multicolumn{3}{c}{8-\htwoo} \\
		& Error & Gates & Depth & Error & Gates & Depth & Error & Gates & Depth \\
		\hline
		A2C         & $7.29 \times 10^{-6}$   & 15  & 8   & $8.25 \times 10^{-5}$   & 26  & 15  & $4.13 \times 10^{-4}$   & 105  & 44  \\
		A3C         & $3.33 \times 10^{-5}$   & 19  & 12  & \bf$6.99 \times 10^{-7}$ & 32  & 14  & $6.87 \times 10^{-4}$   & 85   & 34  \\
		DDQN        & $3.72 \times 10^{-6}$   & 7   & 4   & $5.04 \times 10^{-6}$   & 65  & 35  & $1.82 \times 10^{-4}$   & 173  & 158 \\
		DQN         & \bf$1.11 \times 10^{-6}$ & 17  & 11  & $1.44 \times 10^{-5}$   & 62  & 30  & $2.71 \times 10^{-5}$   & 137  & 68  \\
		DQN\_PER    & $4.57 \times 10^{-6}$   & 8   & 7   & $2.73 \times 10^{-6}$   & 34  & 17  & $5.67 \times 10^{-5}$   & 64   & 32  \\
		DQN\_rank   & $4.99 \times 10^{-6}$   & 24  & 14  & $3.09 \times 10^{-6}$   & 69  & 31  & \bf$3.13 \times 10^{-6}$ & 168  & 105 \\
		Dueling\_DQN& $8.43 \times 10^{-6}$   & \bf2 & \bf2 & $8.92 \times 10^{-6}$ & 28  & 17  & $1.50 \times 10^{-5}$   & 216  & 112 \\
		PPO         & $1.46 \times 10^{-6}$   & 25  & 21  & $1.21 \times 10^{-4}$   & \bf11 & \bf6 & $1.22 \times 10^{-5}$   & 97   & 41  \\
		TPPO        & $3.86 \times 10^{-6}$   & 15  & 11  & $1.01 \times 10^{-5}$   & 26  & 12  & $4.77 \times 10^{-5}$   & \bf34   & \bf21  \\
		\hline
	\end{tabular}
	\caption{\textbf{Benchmarking RL agents for QAS under realistic noise applies 0.1\% single-qubit and 0.01\% two-qubit depolarizing noise after each gate}. Results report the best-performing seeds across runs, highlighting agent robustness and efficiency in hardware-like conditions.}
	\label{tab:noise}
\end{table*}

\section{Results}
\subsection{Noiseless}
The benchmarking study in Fig.~\ref{fig:noiseless_benchmark} systematically evaluates reinforcement learning (RL) algorithms for quantum architecture search (QAS) across three representative variational quantum algorithms: variational quantum state diagonalization (VQSD, 2-qubit), variational quantum classifier (VQC, 3-qubit), and variational quantum eigensolver (VQE, 4-\htwo, 6-\behtwo, 8-\htwoo). The VQE figure highlights that DDQN, DQN and DQN\_rank consistently minimize circuit depth, CX count, and single-qubit gate count as problem size increases, while PPO and TPPO emerge as strong contenders for intermediate and larger problem sizes, often matching or outperforming A3C and DDQN in both depth and gate efficiency. For VQSD and VQC, both value-based (DQN, Dueling DQN) and advanced policy-gradient methods (A3C, TPPO) yield compact, expressive circuits, with robust exploration and stable policy improvement proving advantageous in smaller and moderately complex settings. DDQN and similar variants generally lag in larger VQE problems due to conservative updates. Overall, no single RL paradigm is universally optimal; rather, algorithmic features such as asynchronous updates, trust-region constraints, and action prioritization are critical for hardware-efficient quantum circuits, and the RL algorithm should be selected based on the structure and requirements of the quantum problem.

Notably, \textit{BenchRL-QAS on the 6-$\behtwo$ outperforms TF-QAS~\cite{he2024training}, which achieves an error $0.0018$ with $57$ gates. In contrast, we obtain an error on the order of $10^{-6}$-over three orders of magnitude improvement, while reducing the total gate to $16-22$ depending on the RL-agent}. This improvement in both accuracy and hardware efficiency establishes a new state-of-the-art for VQE ansatz synthesis.

\subsection{Noisy}
The performance results in Table~\ref{tab:noise} demonstrate significant variability among RL algorithms when tasked with quantum circuit design under realistic noise settings. We consider 0.1\% 1-qubit depolarizing noise and 0.01\% of 2-depolarizing noise to benchmark the performance of RL agents. To make the setting more realistic the noise is applied after the application of each gate while constructing the ansatz. Some algorithms, such as \textit{DQN and PPO, achieve notably low error rates for certain qubit sizes, while others, such as DDQN and Dueling\_DQN, excel in minimizing gate counts or circuit depth. This diversity in strengths highlights that no single algorithm consistently outperforms all others across every metric (error, gates, depth) and problem size}.

If we consider the weighted ranking metric in Eq.~\ref{eq:weighted_ranking_metric} with $w_E=0.6$, $w_G=0.1$, $w_D=0.3$, and $w_T=0$ the best RL agents are DDQN (4-\htwo), PPO (6-\behtwo), and TPPO (8-\htwoo). This weighting is well-suited for realistic noisy quantum devices, as it prioritizes accuracy and depth crucial for noise resilience, while still considering gate efficiency.

\begin{table*}[t!]
	\centering
	\fontsize{9}{9}\selectfont
	\begin{tabular}{lccc|cc|c}
		\toprule
		& \multicolumn{3}{c|}{{RL-VQE}} & {RL-VQC} & {RL-VQSD} & {RL-State preparation$^*$} \\
		{RL Algorithm} & {4-\htwo} & {6-\behtwo} & {8-\htwoo} & {3-qubit} & {2-qubit} & {3-qubit} \\
		\midrule
		PPO            & 0.44   & \bf 1.56   & \bf 35.06   & 2.96  & 1.11 & \bf 0.04194 \\
		Dueling\_DQN   & 0.81   & 2.57   & 72.51   & 4.60  & 1.53 & 0.15503 \\
		DDQN           & \bf 0.24   & 2.64   & 66.58   & \bf 0.81  & 1.70 & 0.07143 \\
		A3C            & 0.78   & 2.68   & 56.11   & 8.27  & 2.43 & 0.16253 \\
		DQN            & 0.76   & 2.58   & 66.44   & 4.62  & 1.58 & 0.14595 \\
		DQN\_PER       & 1.03   & 3.11   & 85.55   & 4.76  & 1.85 & 0.23197 \\
		DQN\_rank      & 1.04   & 3.27   & 159.15  & 4.64  & 1.90 & 0.25066 \\
		TPPO           & 0.49   & 1.84   & 132.52  & 1.51  & \bf 0.95 & 0.04844 \\
		A2C            & 0.75   & 2.19   & 265.87  & 2.88  & 1.66 & 0.04729 \\
		\bottomrule
	\end{tabular}
	\caption{\textbf{Average per-episode runtime (in seconds) for each RL algorithm across quantum architecture search tasks (RL-VQE, RL-VQC, RL-VQSD and RL state preparation)}. Lowest times per task are highlighted in green color. $^*$ denotes the action space for this task is non-parameterized.}
	\label{tab:episode_time_stateprep}
\end{table*}

\section{Runtime Analysis of Agents}
\label{appendix:runtime_analysis}

Table~\ref{tab:episode_time_stateprep} presents the average per-episode runtime (in seconds) for RL algorithms across the RL-VQE, RL-VQC, and RL-VQSD tasks, with a final column highlighting 3-qubit state preparation, thereby illustrating each method's computational efficiency. Policy-gradient methods, particularly PPO, consistently achieve the fastest times in most VQE scenarios (1.56s for $6$-\behtwo, 35.06s for $8$-\htwoo) and the lowest state preparation runtime (0.0419s), while TPPO is also efficient (e.g., 1.84s for $6$-\behtwo and 0.95s for 2-qubit VQSD); DDQN excels in $4$-\htwo (0.24s) and is competitive in VQC (0.81s) and state preparation (0.0714s). In contrast, value-based methods tend to incur greater costs as system size grows, exemplified by DQN\_rank’s 159.15s per episode on $8$-\htwoo and generally slower performance from DQN\_PER and A3C, with the slowest recorded runtime from A2C (265.87s on $8$-\htwoo). Notably, state preparation is uniformly swift across all methods (all $<$ 0.26s), suggesting computational bottlenecks arise from agent-environment interaction and neural network training rather than quantum simulation overhead. These results indicate that policy-gradient approaches like PPO and TPPO scale more efficiently and are thus better suited to large-scale quantum circuit design, while value-based methods require further optimization for practical use in large systems.

\subsection{Summary}
Our benchmarking across noiseless and noisy regimes reveals a central result: \textit{no single RL algorithm is universally optimal for quantum circuit design}. This aligns with the No Free Lunch Theorem~\cite{wolpert1997no}, which states that no optimizer consistently outperforms others across all tasks. Empirically, we find that algorithmic performance, measured by circuit depth, gate count, and error which varies with problem structure and noise. Value-based methods (e.g., DQN, DQN\_rank) excel for VQE in noiseless settings, while policy-gradient methods (e.g., A3C, TPPO) are more effective for VQSD and VQC. Under noise, the best-performing algorithm depends on the specific metric and qubit size. \textit{These results highlight that RL algorithm selection must be tailored to each quantum problem and noise regime. Systematic benchmarking, as performed here, is essential for identifying the most suitable approach}. In summary, our study provides practical evidence for the NFL principle in RL-based quantum circuit design: algorithmic strengths are context-dependent, and no single method dominates across all scenarios.

\section{Discussion and Conclusion}
We present \textbf{BenchRL-QAS}, a unified benchmarking framework for reinforcement learning (RL) algorithms in quantum architecture search (QAS), systematically evaluating both policy-gradient and value-based agents across a diverse set of quantum variational tasks and system sizes (2- to 8-qubit). BenchRL-QAS introduces a weighted performance ranking for fair comparison and makes all code and data publicly available for reproducibility. Our results demonstrate that no single RL agent is universally optimal which is a direct reflection of the ``No Free Lunch'' principle~\cite{wolpert1997no}: value-based methods (e.g., DQN, DQN rank, Dueling DQN) excel in deeper VQE tasks, while policy-gradient methods (e.g., A3C, TPPO) are more effective for structured or smaller problems like VQSD and VQC; under noise, optimal performance is metric- and problem-dependent. This highlights the need for tailored algorithm selection and comprehensive benchmarking in RL-driven quantum circuit design.

\paragraph{Limitations and future work.}
While BenchRL-QAS represents the most comprehensive RL-QAS benchmarking effort to date, several limitations remain. Our current study evaluates a utilizes vanilla curriculum RL-QAS, which is not the best performing if we consider advanced methods such as TensorRL-QAS~\cite{kundu2025tensorrl} or Gadget RL-QAS~\cite{kundu2025reinforcementlearninggadget}. Moreover, while the tasks in this study span multiple VQA classes such as quantum approximate optimization algorithm \cite{farhi2014quantumapproximateoptimizationalgorithm} were not included.
Additionally, the noisy scenario was limited to the VQE algorithm. Incorporating a broader set of quantum optimization problems under realistic QPU constraints (e.g., qubit connectivity and/or gate fidelity) would enhance the generality of the framework.

\section{Acknowledgments}
We thank QOSF for providing the platform for collaboration, we also thank Maria Demidik for fruitful discussions. Finally we thank Center of Science (CSC) IT centre in Finland for providing us with the CPU and GPU resources.

\bibliography{aaai25}

\end{document}